\documentclass[aps,prb,reprint,shortbibliography,superscriptaddress]{revtex4-1}
\usepackage{ragged2e}
\usepackage{setspace}
\usepackage{amsmath}
\usepackage{graphicx}
\usepackage[nearskip,margin = 0pt]{subfig}
\usepackage{subfig}

\usepackage{verbatim}
\usepackage{amsfonts}
\usepackage{amssymb}
\usepackage{epstopdf} 
\usepackage{xcolor}
\DeclareGraphicsExtensions{.pdf,.eps,.png,.jpg,.mps} 
\usepackage{hyperref}
\hypersetup{
    colorlinks=true,
    linkcolor=blue,
    citecolor=blue,
    filecolor=blue,      
    urlcolor=blue,
}
\begin{document}
\title{Nanometric dual-comb ranging using photon-level microcavity solitons}

\author{Zihao Wang}
\affiliation{State Key Laboratory of Precision Measurement Technology and Instruments, Department of Precision Instruments, Tsinghua University, Beijing 100084, China}

\author{Yifei Wang}
\affiliation{State Key Laboratory of Precision Measurement Technology and Instruments, Department of Precision Instruments, Tsinghua University, Beijing 100084, China}

\author{Baoqi Shi}
\affiliation{International Quantum Academy, Shenzhen 518048, China}
\affiliation{Department of Optics and Optical Engineering, University of Science and Technology of China, Hefei, Anhui 230026, China}

\author{Wei Sun}
\affiliation{International Quantum Academy, Shenzhen 518048, China}

\author{Changxi Yang}
\affiliation{State Key Laboratory of Precision Measurement Technology and Instruments, Department of Precision Instruments, Tsinghua University, Beijing 100084, China}

\author{Junqiu Liu}
\affiliation{International Quantum Academy, Shenzhen 518048, China}
\affiliation{Hefei National Laboratory, University of Science and Technology of China, Hefei 230088, China}

\author{Chengying Bao}
\email{cbao@tsinghua.edu.cn}
\affiliation{State Key Laboratory of Precision Measurement Technology and Instruments, Department of Precision Instruments, Tsinghua University, Beijing 100084, China}
\begin{abstract}
Absolute distance measurement with low return power, fast measurement speed, high precision, and immunity to intensity fluctuations is highly demanded in nanotechnology. However, achieving all these objectives simultaneously remains a significant challenge for miniaturized systems. 
Here, we demonstrate dual-comb ranging (DCR) that encompasses all these capabilities by using counter-propagating (CP) solitons generated in an integrated Si$_3$N$_4$ microresonator. We derive equations linking the DCR precision with comb line powers, revealing the advantage of microcomb's large line spacing in precise ranging. Leveraging the advantage, our system reaches 1-nm-precision and measures nm-scale vibration at frequencies up to 0.9 MHz. We also show that precise DCR is possible even in the presence of strong intensity noise and loss, using a mean received photon number as low as 5.5$\times$10$^{-4}$ per pulse. 
Our work establishes an optimization principle for dual-comb systems and bridges high performance ranging with foundry-manufactured photonic chips.
\end{abstract}

\maketitle

\noindent \textbf{Introduction.} Nanotechnology is built on nanometrology. Dual-comb ranging (DCR), which uses a pair of coherent optical frequency combs for measurement, provides an elegant approach to reach nanometre-precision with fast measurement speed \cite{kim_NP2009combs,Newbury_NP2009rapid,Wu_Engineering2018,Newbury_Nature2022time}. 
The advent of optical frequency microcombs \cite{Kippenberg_NP2014,Kippenberg_Science2018Review,Diddams_Science2020optical,Bowers_NP2022integrated} has brought photonic integrated circuit (PIC)-based DCR system within reach \cite{Kippenberg_Science2018Range,Vahala_Science2018Range,yang2024optical}. PICs can be processed on wafer-scale, which can greatly advance the widespread adoption of nanometric distance sensors in nanotechnology.  
Moreover, nanophotonic microcombs further offer unique advantages including large line spacing ($f_r$), large repetition rate difference ($\delta f_r$), high single comb line power and passive mutual coherence \cite{Kippenberg_Science2018Review}. These are particularly useful to enhance the DCR speed and precision and to enable DCR with low return power. However, the potential of microcomb-based DCR has not been fully released. For example, microcomb-based DCR has reached an update rate of 100 MHz and a precision of 12 nm \cite{Kippenberg_Science2018Range}, but further precision improvement to 1-nm-scale has been hampered by a lack of mutual coherence. The mutual coherence can be derived by generating counter-propagating (CP) solitons to have Vernier-like frequency locking (VFL) \cite{Vahala_NP2017Counter,Vahala_Science2019Vernier,Vahala_NC2021DCS,Vahala_Science2018Range,Bao_PRAppl2023vernier,Bao_arXiv2024rhythmic}. However, the previously reported silica CP soliton-based DCR system did not use a VFL state, resulting in limited precision and update rate of 200 nm and 6 kHz, respectively \cite{Vahala_Science2018Range}. More generally, DCR lacks a quantitative model to guide the optimization of ranging systems, despite the tremendous progression in the last decades. 

Here, we use mutually coherent CP solitons generated in a foundry-manufactured Si$_3$N$_4$ microresonator \cite{Liu_PR2023foundry,Bao_arXiv2024rhythmic} for DCR and reach the 1-nm-precision regime. We derive and verify equations establishing the relationship between DCR precision and spectral signal-to-noise ratio in the radiofrequency (RF) domain (denoted as rSNR). It not only fills the longstanding theoretical understanding gap for DCR, but also impacts precise spectroscopy using dual-comb phase measurement \cite{Newbury_Optica2016}. 
The coherent soliton pair also enables precise DCR with a received power as low as 7 pW or 5.5$\times$10$^{-4}$ photon per pulse for our 100 GHz soliton stream. The power used for photodetection is more than 60 dB lower than reported microcomb-based ranging systems \cite{Kippenberg_Science2018Range,Vahala_Science2018Range,Kippenberg_Nature2020massively,Zhang_PR2020long,Wong_PRL2021nanometric,Kippenberg_NP2023chaotic,Wang_NP2023breaking}, 
and the photon number is an order of magnitude lower than the time-programmable frequency combs DCR system \cite{Newbury_Nature2022time}. 
Our system's reliance on the relative phase stability of CP solitons makes it immune to intensity fluctuations, setting it apart from other high-precision ranging techniques including the electro-optical sampling (EOS) \cite{Kim_NP2020ultrafast}, time-of-flight (ToF) \cite{Vahala_Science2018Range}, chaotic ranging \cite{Wang_NP2023breaking,Kippenberg_NP2023chaotic} and dispersive interferometry \cite{Zhang_PR2020long,Wong_PRL2021nanometric} techniques. In addition, we perform DCR at a rate of 1.83 MHz without optical amplification and measure fast vibrations up to 900 kHz with a sensitivity of 0.4 nm/$\sqrt{\rm Hz}$. The fast dual-comb vibrometry (DCV) is possible due to the high power per comb line in microcombs, although low total power has been perceived as a significant drawback for soliton microcombs. Our work highlights the advantages of microcombs for precise ranging, and can be used in surface profiling of samples with low reflection (e.g., integrated circuits and thin films for PICs) \cite{Kim_NP2020ultrafast,joo2013femtosecond}, strain sensing \cite{Kim_NP2020ultrafast,gagliardi2010probing}, and human voice recovery \cite{Boudreau_OE2014range}. 

\begin{figure*}[th!]
\begin{centering}
\includegraphics[width=\linewidth]{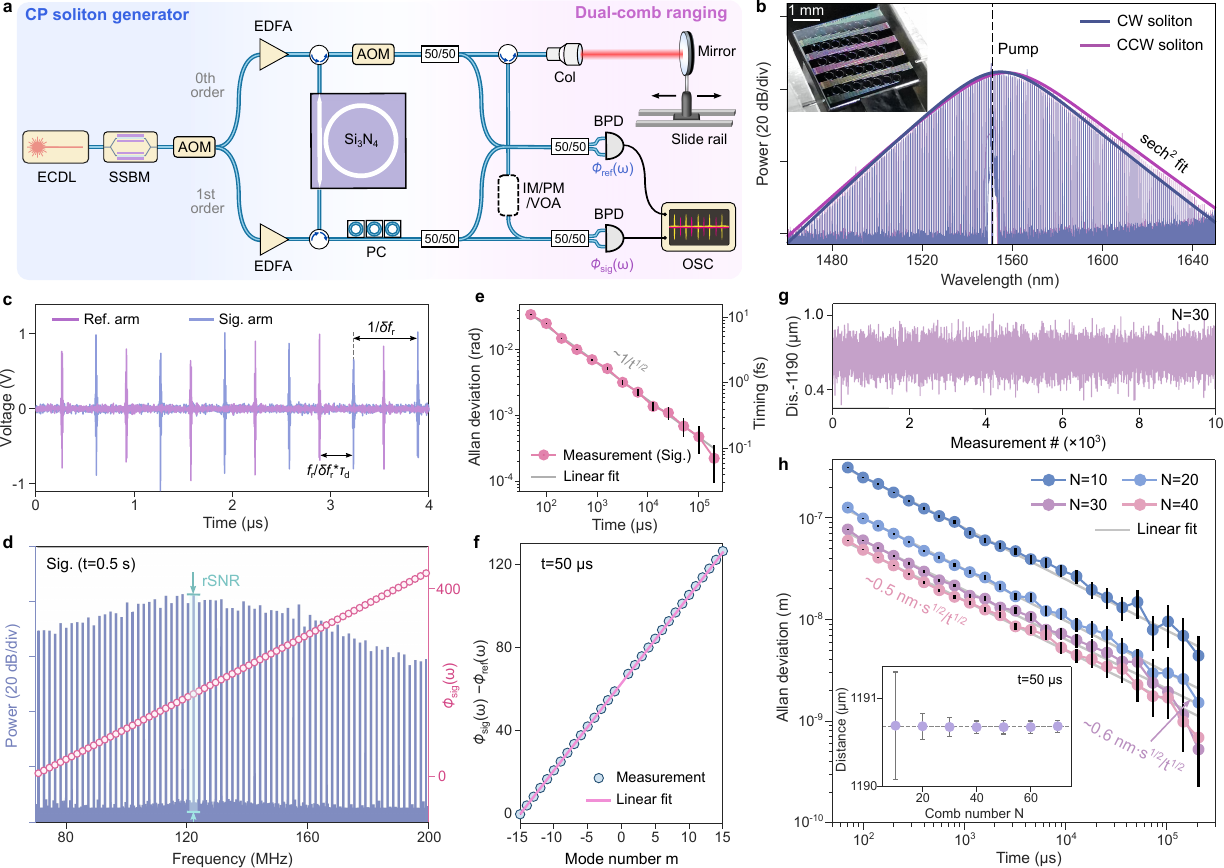}
\captionsetup{singlelinecheck=off, justification = RaggedRight}
\caption{{\bf Coherent dual-comb ranging (DCR) using counter-propagating (CP) solitons}
\textbf{a,} Experimental setup for the CP solitons generation and DCR. ECDL, external cavity diode laser; SSBM, single-sideband modulator; AOM, acousto-optical modulator; FBG, fibre Bragg grating; EDFA, erbium-doped fibre amplifier; PC, polarization controller; Col, collimator; BPD, balanced photodetector.
\textbf{b,} Optical spectra of the CP solitons. The inset shows the nanophotonic chip.
\textbf{c,} Dual-comb inteferogram in the signal and reference arms.
\textbf{d,} Power spectrum of the signal arm and the corresponding phase spectrum.
\textbf{e,} Allan deviation of phase different between the 4th and the 9th lines. 
\textbf{f,} Phase difference between the signal and the reference arms measured in 50 $\mu$s. 
\textbf{g,} Fitting the phase difference in panel \textbf{f} yields the distance over 10$^4$ time slots.
\textbf{h,} DCR Allan deviation when selecting different number of comb lines for fitting, all showing $t^{-1/2}$ scaling. The inset shows all the selections yield the same distance, but with different standard deviation (see error bars).}
\label{fig1}
\end{centering}
\end{figure*}

\begin{figure*}[th!]
\begin{centering}
\includegraphics[width=\linewidth]{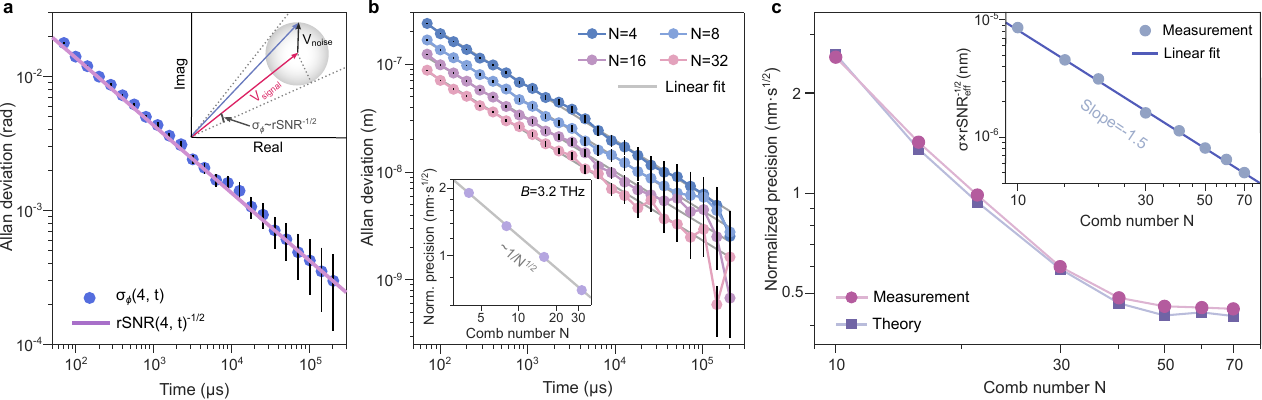}
\captionsetup{singlelinecheck=off, justification = RaggedRight}
\caption{{\bf Measurement precision of DCR.} \textbf{a,} Allan deviation of the phase of the 4th comb line, which equals 1/$\sqrt{\rm rSNR}$. The inset shows an illustration of the relationship between rSNR and phase deviation. 
\textbf{b,} DCR precision for a fixed comb bandwidth $B$=3.2 THz, but selecting comb lines with different spacing (thus, different used comb line number $N$). The inset shows the normalized precision improves as $N^{-1/2}$.
\textbf{c,} Measured DCR precision agrees with the theoretical precision determined by Eq. \ref{eq2Sigma}. The inset confirms the $N^{-3/2}$ trend in our theory.}
\label{fig1A}
\end{centering}
\end{figure*}

\begin{figure*}[th!]
\begin{centering}
\includegraphics[width=\linewidth]{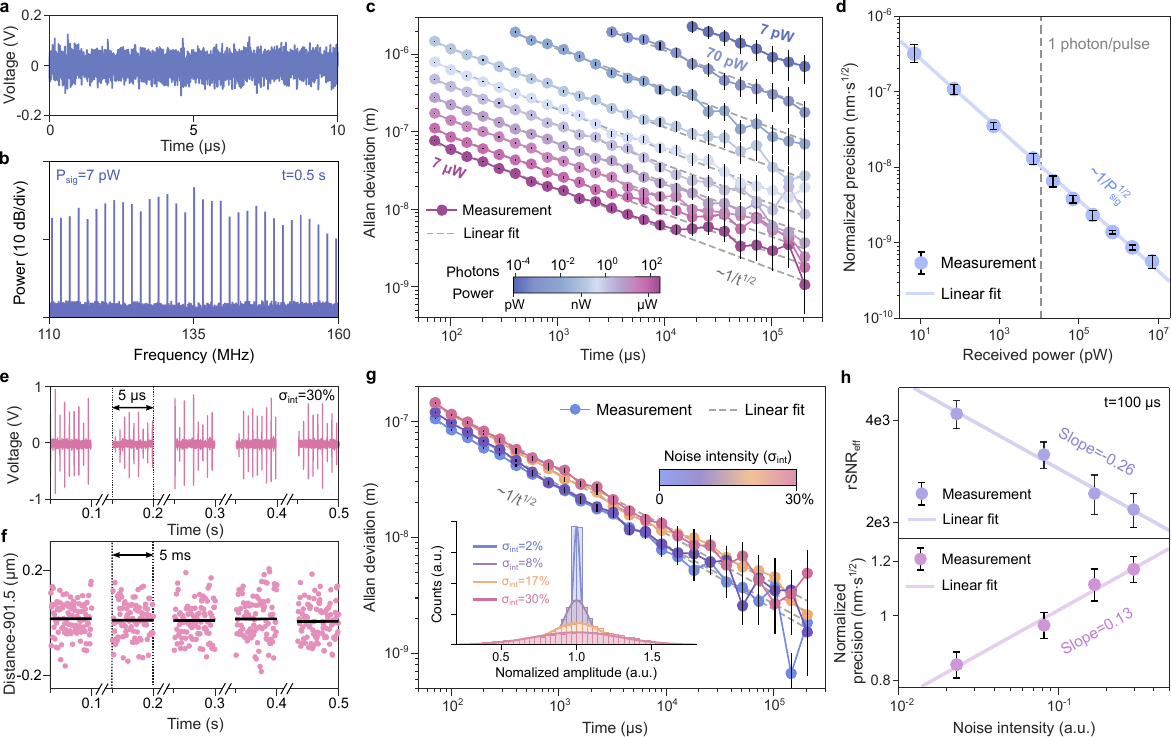}
\captionsetup{singlelinecheck=off, justification = RaggedRight}
\caption{ {\bf DCR with low received power and strong intensity noise.}
\textbf{a,} Dual-comb interferogram signal with a received power of 7 pW. 
\textbf{b,} RF spectrum of the signal in panel \textbf{a} after $t$=0.5 s coherent averaging. 
\textbf{c,} DCR Allan deviation of a series of received powers, all exhibiting $t^{-1/2}$ scaling.
\textbf{d,} Normalized DCR precision, showing an inverse square-root relationship with the received power.
\textbf{e,} Measured RF pulses have randomly fluctuating amplitudes when introducing intensity noise on the received microcomb.
\textbf{f,} Measured distance in five 5 ms slots separated by 0.1 s (50 $\mu$s per measurement point) and solid lines are the average distance. 
\textbf{g,} DCR Allan deviation under different intensity noise. The inset shows the distribution of the RF pulse amplitude. In the absence of intensity noise, the amplitude has a 2\% fluctuation. The added intensity noise can introduce 30\% amplitude fluctuation. 
\textbf{h,} Deterioration of the rSNR under different intensity noise, which results in the increase of the DCR Allan deviation.}
\label{fig2}
\end{centering}
\end{figure*}

\noindent \textbf{Mutually coherent DCR with nm-precision.} The interferometric DCR principle is rooted in the Fourier transform relationship, that a temporal shift $\tau_d$ corresponds to a linear phase of $\omega\tau_d$ in the frequency domain ($\omega=m\omega_r$, where $m$ is the comb line number with respect to the pump and $\omega_r$ is the angular repetition rate). By measuring the multi-heterodyne phases for the signal and reference arms, denoted as $\phi_{\rm sig}(\omega)$ and $\phi_{\rm ref}(\omega)$ in Fig. \ref{fig1}a, we can derive $\tau_d$ and distance $L$ as,
\begin{equation}
L=c\tau_d/2=c(\phi_{\rm sig}(\omega)-\phi_{\rm ref}(\omega))/2\omega.
\label{eq1}
\end{equation}
Hence, high mutual coherence and stable $\phi_{\rm sig}, ~\phi_{\rm ref}$ are the keys for precise DCR. Note that we omit the influence of air group index and treat the pulse group velocity as the vacuum light velocity $c$. This group index impact can be included by using two-colour measurements \cite{Minoshima_AO2000}.

Our setup to generate CP solitons with VFL is shown in Fig. \ref{fig1}a, see Methods. The generated soliton microcombs (Fig. \ref{fig1}b) were heterodyne beat on balanced photodetectors (BPDs) to generate the DCR signals. An example of the interferogram with $\delta f_r$=1.62 MHz is shown in Fig. \ref{fig1}c. We then Fourier transform the RF pulses to have the power spectrum and $\phi_{\rm sig}(\omega)$ shown in Fig. \ref{fig1}d. A power rSNR (signal over average noise floor) exceeding 80 dB can be obtained within a measurement time $t$=0.5 s. To show the phase stability between the CP solitons, we plot the Allan deviation of $\phi_{\rm sig}(9 \omega_r)-\phi_{\rm sig}(4 \omega_r)$ in Fig. \ref{fig1}e, which scales as $t^{-1/2}$ and reaches 4.5 mrad at 0.1 s. The relative timing stability between CP solitons can be estimated as $(\phi_{\rm sig}(9 \omega_r)-\phi_{\rm sig}(4 \omega_r))/5\omega_r$, reaching 0.1 fs at 0.1 s.

Then, we analyze $\phi_{\rm sig}-\phi_{\rm ref}$ using RF pulses measured within 50 $\mu$s and observe an excellent linearity for the used $N$=30 lines (excluding the pump, see Fig. \ref{fig1}f). Fitting the relative phase yields the distance in multiple 50 $\mu$s slots (Fig. \ref{fig1}g). The DCR precision is evaluated by Allan deviation and shows a $t^{-1/2}$ scaling with a normalized precision of 0.6 nm$\cdot\sqrt{\rm s}$ (Fig. \ref{fig1}h). 
We further changed the used comb line number from $N$=30 to other numbers to evaluate the precision. Normalized precision as high as 0.5 nm$\cdot\sqrt{\rm s}$ (reaching 1.1-nm-precision at 0.2 s) is possible for our system. The inset of Fig. \ref{fig1}h confirms that our measurement always yields the same distance when using different comb line number $N$. 

\begin{figure*}[th!]
\begin{centering}
\includegraphics[width=\linewidth]{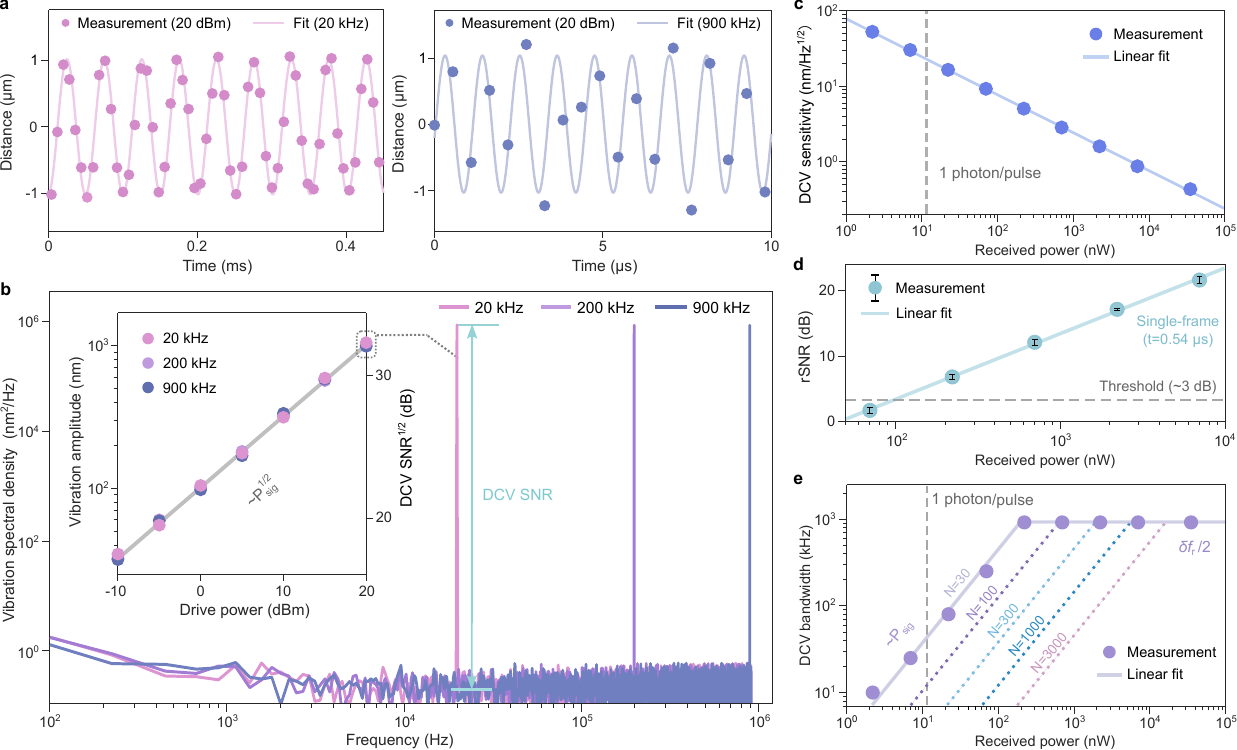}
\captionsetup{singlelinecheck=off, justification = RaggedRight}
\caption{{\bf Near Megahertz dual-comb vibration (DCV) measurements.}
\textbf{a,} Measured distance change with drive frequencies of 20 or 900 kHz. 
\textbf{b,} DCV spectra  measured at three drive frequencies, all yielding sharp peaks with a high signal-to-noise ratio (SNR). The inset shows the measured vibration amplitude versus the drive power of the phase modulator (PM). 
\textbf{c,} The DCV sensitivity (determined by the noise floor in panel \textbf{b}) scales in a square-root way with the received microcomb power. 
\textbf{d,} Deduced rSNR for a single-frame inteferogram under different received powers. This rSNR should exceed 3 dB for a reliable DCV measurement.
\textbf{e,} The highest vibration frequency that can be measured with different received powers. Below 200 nW power, the highest frequency decreases linearly with the power. When the comb line number becomes larger for a fixed total power, DCV at $\delta f_r$/2 will need a higher received power.
}
\label{fig3}
\end{centering}
\end{figure*}

\begin{figure}[t]
\captionsetup{singlelinecheck=no, justification = RaggedRight}
\includegraphics[width=0.98\linewidth]{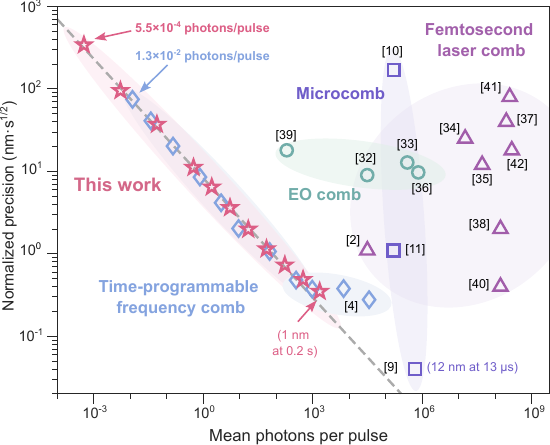}
\caption{
{\bf Performance comparison with other dual-comb ranging systems.} Our system feature low received pulse energy and high precision. EO comb, electro-optical comb.}
\label{fig5}
\end{figure}


\noindent \textbf{Precision of DCR.}  Theoretically, the DCR precision is derived as (see Methods),

\begin{equation}
\sigma = \frac{\sqrt{3}\sqrt{1+\alpha}c}{N \omega_r \sqrt{N}\sqrt{\rm rSNR_{\rm eff}}} = \frac{\sqrt{3}\sqrt{1+\alpha}c}{2\pi B \sqrt{N}\sqrt{\rm rSNR_{\rm eff}}},
\label{eq2Sigma}
\end{equation}
where rSNR$_{\rm eff}$ is the effective rSNR for the signal arm,  
$B$=$N$$f_r$ is the used comb bandwidth, and $\alpha$ is the ratio between the rSNR$_{\rm eff}$ in the reference and the signal arms. The phase of a measured RF tone has contribution from both signal and noise (inset of Fig. \ref{fig1A}a). Thus, the phase deviation of a measured RF line ($\sigma_\phi$) is determined by the rSNR determines as, 

\begin{equation}
\sigma_\phi(m,t) = \frac{1}{\sqrt{{\rm rSNR}(m,t)}}.
\label{eq2ASigmaPhi}
\end{equation}
$\sigma_\phi(4,t)$ plotted in Fig. \ref{fig1A}a confirms this relationship. $\sigma_\phi$ divided by 2$\pi B$ yields the timing (thus, distance) deviation. The $1/\sqrt{N}$ term is added as fitting multiple points leads to higher stability \cite{montgomery2021introduction} (Supplementary Sec. 1). To verify it, we analyzed the DCR precision for a fixed $B$=3.2 THz but selecting lines with different spacings (thus, different $N$). A $1/\sqrt{N}$ scaling is observed for the normalized precision (Fig. \ref{fig1A}b). Finally, $\alpha$ is included as the DCR signal is derived from the phase difference between two arms (Eq. \ref{eq1}). 

Our measured DCR precision is in excellent agreement with the theory (Fig. \ref{fig1A}c). The precision no longer improves evidently for $N$$>$40. This is because rSNR($m$) decreases for large $|m|$, resulting in a reduced rSNR$_{\rm eff}$. We numerically calculated rSNR$_{\rm eff}$ based on rSNR($m$) of the used microcomb lines (Methods and Supplementary Fig. S1). After taking the decrease of rSNR$_{\rm eff}$ into account, a scaling of $N^{-3/2}$ is observed (see Eq. \ref{eq2Sigma} and inset of Fig. \ref{fig1A}b). 

Therefore, the high DCR precision in our system can be attributed to the high rSNR and broad usable bandwidth. rSNR$_{\rm eff}$ can be further written as,

\begin{equation}
{\rm rSNR_{\rm eff}}=\frac{K(B)}{N^2} {\frac{P_{\rm LO}P_{\rm sig}}{S_{\rm n} f_{\rm BW}}}=\frac{K(B)}{N^2} \frac{f_r^2E_{\rm LO}E_{\rm sig}t}{S_{\rm n}},
\label{eq2rSNR}
\end{equation}
where $S_{\rm n}$ is the spectral power density of the RF noise floor, and $f_{\rm BW}$=1/$t$ is the resolution bandwidth; $P_{\rm sig(LO)}$ and $E_{\rm sig(LO)}$ are the received signal (local) optical comb power and energy, respectively; $K(B)$ is a conversion coefficient that includes response from the BPD and variation of rSNR($m$) within the used bandwidth. By combining Eq. \ref{eq2Sigma} and Eq. \ref{eq2rSNR}, it can be seen that $\sigma$ is proportional to $\sqrt{N}$ for a given bandwidth $B$ and comb powers. Note that the comb power is ultimately limited by the saturation of the BPDs. Thus, small comb line numbers for microcombs are beneficial for enhancing DCR precision.



\noindent \textbf{DCR with intensity loss or noise.} Due to small $N$, only low comb power and pulse energy are needed for precise DCR when using microcombs (Eqs. \ref{eq2Sigma}, \ref{eq2rSNR}).  
In experiments, we adjusted the loss in the signal arm (Fig. \ref{fig1}a) to measure DCR precision under different received powers. When the power was reduced to 7 pW (attenuated by 67 dB), the dual-comb interferogram is already buried in the noise (Fig. \ref{fig2}a). However, an RF comb with rSNR over 10 dB can still be obtained after coherent averaging 0.5 s data (Fig. \ref{fig2}b). Mutual coherence for VFL solitons is the key to coherent averaging and resolving these RF tones. 

The DCR Allan deviation still scales as $t^{-1/2}$, but with a lower normalized precision of 300 nm$\cdot\sqrt{\rm s}$ (top curve in Fig. \ref{fig2}c). It means a micron-precision can be obtained in 0.1 s using only 5.5$\times$10$^{-4}$ photon per pulse. 
The Allan deviation for a series of received powers is plotted in Fig. \ref{fig2}c, all exhibiting the $t^{-1/2}$ scaling. The normalized DCR precision decreases with the received power in a square-root trend, consistent with Eqs. \ref{eq2Sigma}, \ref{eq2rSNR} (Fig. \ref{fig2}d).

Since interferometric DCR relies upon optical phase measurements, our system is immune against intensity fluctuations. To showcase this feature, we inserted an intensity modulator (IM) and drove it by a broadband noise in the signal arm (Fig. \ref{fig1}a). The measured RF pulses have strong fluctuations in the amplitude (Fig. \ref{fig2}e). Despite these fluctuations, the measured average distance remains the same for five 5 ms slots separated by 0.1 s ($t$=50 $\mu$s for a single data point, see Fig. \ref{fig2}f). 
The distribution of the RF pulse amplitudes subject to different intensity noise levels is shown in the inset of Fig. \ref{fig2}g (the blue one is the case without IM noise). The added intensity noise can cause the RF pulses to have a 30\% intensity fluctuation (see the labelled standard deviation $\sigma_{\rm int}$). DCR Allan deviation retains the $t^{-1/2}$ scaling in the presence of the intensity noise (Fig. \ref{fig2}g, average received power is 15 $\mu$W). Nanometre-scale precision is still possible with $\sigma_{\rm int}$=30\%. 

The Allan deviation does increase with $\sigma_{\rm int}$, as the intensity noise deteriorates rSNR. rSNR$_{\rm eff}$ for the used 30 lines decreases with a slope of $-$0.26 in a log-log plot versus $\sigma_{\rm int}$, while the change of the DCR Allan deviation has a slope of 0.13 (Fig. \ref{fig2}h). Such a $-$0.5 relationship between them further strengthens our theory in Eq. \ref{eq2Sigma}. Note that the rSNR deterioration does not impact the DCR measured distance (Fig. \ref{fig2}f); it only means a longer measurement time is needed to reach a desired precision. The intensity noise immunity distinguishes interferometric DCR from other comb-based ranging techniques, that rely upon power detection \cite{Kim_NP2020ultrafast,Vahala_Science2018Range,Zhang_PR2020long,Wong_PRL2021nanometric,Wang_NP2023breaking,Kippenberg_NP2023chaotic}. The immunity can be quite useful when the measured target surface has varying reflection coefficients and may be used to distinguish amplitude and phase fluctuations induced by air turbulence \cite{Newbury_PRL2015broadband}.

\noindent \textbf{Near Megahertz DCV.} The high rSNR also enables DCR using a single-frame interferogram at a rate of $\delta f_r$=1.83 MHz, i.e., DCV at a rate up to $\delta f_r$/2. Similar single-frame DCR was demonstrated in ref. \cite{Kippenberg_Science2018Range}, but optical power amplification was used. Here, we implement it without optical amplification. We experimentally inserted a phase-modulator (PM) into the signal arm to introduce optical path length change (Fig. \ref{fig1}a). Figure \ref{fig3}a shows the measured optical path length change with a $\sim$1 $\mu$m amplitude at update rates of 125 and 900 kHz (close to $\delta f_r$/2), when driving the PM at frequencies of 20 and 900 kHz, respectively. 
In the frequency domain, the measured distance changes correspond to sharp peaks at 20, 200 and 900 kHz as shown in Fig. \ref{fig3}b (drive frequency at 200 kHz was also measured). The corresponding peak intensities are the same for all the three drive frequencies and are about 66 dB higher than the average noise floor. When varying the drive power for the PM, we observed the peak amplitude scales in a square-root way with the drive power for all the three drive frequencies (inset of Fig. \ref{fig3}a). Such a relationship verifies the high accuracy of our DCV measurement.

The noise floor in Fig. \ref{fig3}b provides a measure of the DCV sensitivity. We summarize this sensitivity versus received powers in Fig. \ref{fig3}c. For a nearly white noise floor, its noise spectral amplitude density has a linear relationship with the Allan deviation in DCR \cite{Barnes1971characterization}. Since the DCR precision has an inverse square-root relationship with the received power (Fig. \ref{fig2}d and Eqs. \ref{eq2Sigma}, \ref{eq2rSNR}), the measured DCV sensitivity also follows an inverse square-root scaling with the received power, and reaches a sensitivity of 0.4 nm/$\sqrt{\rm Hz}$ (Fig. \ref{fig3}c). 

To measure vibration at $\delta f_r$/2, the received power should guarantee a sufficient rSNR($m$, 1/$\delta f_r$) for single-interferogram DCR. It is technically challenging to quantify rSNR($m$, 1/$\delta f_r$) directly, as $f_{\rm BW}$=$\delta f_r$ for a single-interferogram measurement. In practice, we derive it from rSNR at $t$=100 $\mu$s, using the relationship that rSNR scales linearly with $t$ for VFL solitons \cite{Bao_arXiv2024rhythmic}. The minimum rSNR($m$, 1/$\delta f_r$) among the used 30 lines under different received powers is plotted in Fig. \ref{fig3}d. Our data process practice suggests that the minimum rSNR($m$, 1/$\delta f_r$) should exceed 3 dB to enable a reliable measurement. 
The received power needed to reach this threshold is about 100 nW (about 10 photon per pulse). For lower powers, we need to average longer time to reach the required rSNR, and the needed time $t$ increases linearly with decreasing $P_{\rm sig}$. Therefore, the highest vibration frequency decreases linearly with the received power for  $P_{\rm sig}<100$ nW (Fig. \ref{fig3}e). For the current system, tens of kHz vibration can be measured with nW-level received power (over 40 dB loss in the signal arm) without any microcomb amplification. Based on Eq. \ref{eq2rSNR}, the highest measurable DCV frequency decreases linearly with $N$ for a given received microcomb power (Fig. \ref{fig3}e). 

\noindent \textbf{Discussions.}  We compare the system performance with other DCR systems on received pulse energy and normalized precision in Fig. \ref{fig5}. Consistent with Eqs. \ref{eq2Sigma}, \ref{eq2rSNR}, our 100 GHz CP soliton microcomb-based system reaches a high DCR precision using a much lower pulse energy than most other reports \cite{Newbury_NP2009rapid,Kippenberg_Science2018Range,Vahala_Science2018Range,martin2022performance,yang2024optical,weimann2017silicon,liu2011sub,camenzind2022dynamic,zhao2018absolute,kim2020absolute,zhou2019simplified,weimann2018fast,zhu2018synthetic,zhang2014reliable,Newbury_Nature2022time,nurnberg2021dual}. The recently demonstrated time-programmable comb-based DCR systems \cite{Newbury_Nature2022time} has a similar precision using a similar photon consumption, but needs a much more complicated system. Ref. \cite{Kippenberg_Science2018Range} suggests higher `normalized' precision can be possible using more powerful microcombs. CP solitons enable a much higher `absolute' precision (labelled in Fig. \ref{fig5}) by endowing mutual coherence that is missing in ref. \cite{Kippenberg_Science2018Range}.

However, as a characteristics of rhythmically interacting CP solitons \cite{Bao_arXiv2024rhythmic}, a systematic temporal motion needs to be considered and calibrated when measuring over the non-ambiguity range (Supplementary Sec. 2). This soliton motion does not impact the DCR measurement strongly when the measurement range is below micron-scale (Fig. \ref{fig3}b and Supplementary Fig. S2). 

In conclusion, we have demonstrated rapid, intensity noise immune, nanometric absolute ranging using photon-level received soliton pulses. The simplified DCR system tolerates 67 dB power loss to retain the precise ranging capability without using any optical amplification. Mutual coherence between CP solitons is essential to achieve this performance. The simplified DCR system can be used monitor the operation condition of nanodevices. Our theory and measurements elucidate why microcombs fit for dual-comb applications. The theory also provides an invaluable tool to optimize DCR and dual-comb phase spectroscopy \cite{Newbury_PRL2015broadband,Newbury_Optica2016} for all types of comb systems. Recent progress of microcombs bodes well for the full photonic integration of our DCR system \cite{Bowers_Science2021laser}, which could enable nanometric LiDAR on compact, mass-produced nanophotonic chips. 


\vspace{3 mm}
\noindent \textbf{Methods}

{\small
\noindent \textbf{Microresonator and dual-comb generation.} The soliton generation method is the same with ref. \cite{Bao_arXiv2024rhythmic}. A pump laser was split into two arms by an acousto-optical modulator (AOM) with a frequency shift of $\delta\nu_{\rm P}$=55 MHz to counter-pump a high-Q Si$_3$N$_4$ microresonator \cite{Liu_PR2023foundry}. The second AOM in Fig. \ref{fig1}a was used to shift one of the combs by 80 MHz to avoid spectral aliasing \cite{Vahala_NC2021DCS}. In VFL, $\delta f_r$=$\delta\nu_{\rm P}/n$ ($n$ is an integer), which means both the relative repetition frequency and the relative offset frequency are stabilized with RF stability \cite{Vahala_NP2017Counter}. The microresonator is undercoupled with intrinsic and loaded Q-factors being 12.7$\times$10$^6$ and 7.8$\times$10$^6$, respectively. The the local oscillator power on the BPDs was about 40 $\mu$W.

\vspace{1mm}
\noindent \textbf{rSNR$_{\rm eff}$ and derivation of Eq. \ref{eq2Sigma}.} rSNR$_{\rm eff}$ is defined to have a single rSNR for all the used comb lines that leads to the same distance deviation with the measured case which has varying $\text{rSNR}(m)$ for used comb lines. To do it, we first calculate $\text{rSNR}(m)$ for all the used lines within [$-N/2$, $N/2$] (excluding $m$=0, i.e., the filtered pump line). We assume the unperturbed phase has a perfect linear relationship that is $\phi(m)=km$, where $k$ is a slope. Due to noise in the measured RF tones, $\phi(m)$ has an additional noise $\phi_n(m)$ with a deviation of $\sigma_\phi(m)={\rm rSNR}(m)^{-1/2}$ (Eq. \ref{eq2ASigmaPhi}). Then, the fitted slope fluctuates around $k$. Based on rSNR($m$), we numerically generate random phase $\phi_n(m)$ and add it to $km$ to calculate the perturbed slope. After many iterations, the deviation of the fitted slope $\sigma_k$ is calculated. Finally, rSNR$_{\rm eff}$ is determined by having the same slope deviation $\sigma_k$, when assuming all the RF tones have an identical phase deviation of ${\rm rSNR_{eff}}^{-1/2}$.

According to the linear regression theory \cite{montgomery2021introduction}, the slope deviation is $\sigma_k=2\sqrt{3}/({\rm rSNR_{eff}^{1/2}}N^{3/2})$ in the homoscedasticity case (i.e., all the RF tones have the same phase deviation), when $N$ is large (see Supplementary Sec. 2). Thus, Eq. \ref{eq2Sigma} can be derived by swapping dimensionless $m$ to $m\omega_r$. 



\vspace{1 mm}
\noindent \textbf{Data Availability.}
The data that supports the plots within this paper and other findings are available.

\noindent \textbf{Acknowledgements.}
We thank Prof. Yidong Tan at Tsinghua University for discussions and equipment loan. 
The silicon nitride chip used in this work was fabricated by Qaleido Photonics. This work is supported by the National Key R\&D Program of China (2021YFB2801200), by the National Natural Science Foundation of China (62250071, 62175127, 62375150), by the Tsinghua-Toyota Joint Research Fund, and by the Tsinghua University Initiative Scientific Research Program (20221080069). J.L. acknowledges support from the National Natural Science Foundation of China (12261131503), Innovation Program for Quantum Science and Technology (2023ZD0301500), Shenzhen-Hong Kong Cooperation Zone for Technology and Innovation (HZQB-KCZYB2020050), and Shenzhen Science and Technology Program (RCJC20231211090042078).

\vspace{1 mm}

\noindent\textbf{Author Contributions.} Z.W. ran the experiments and analyzed the results with assistance from Y.W., and C.Y.; B.S., W.S. and J.L. prepared and characterized the Si$_3$N$_4$ chip. The project was supervised by C.B.

\vspace{1 mm}
\noindent \textbf{Competing Interests.} The authors declare no competing interests.
}

\bibliography{main}
\end{document}